\documentclass[aps,prl,twocolumn,groupedaddress,amsmath,amssymb]{revtex4-1}
\usepackage{graphicx}  
\usepackage{dcolumn}   

\begin{document}

\title{Experimental demonstration of high-dimensional photonic spatial entanglement between multi-core optical fibers}
\author{Hee Jung Lee}
\author{Sang-Kyung Choi}
\author{Hee Su Park}
\email[Corresponding author. ]{hspark@kriss.re.kr}
\affiliation{Korea Research Institute of Standards and Science (KRISS), Daejeon 34113, South Korea}
\date{\today}\

\begin{abstract}
Fiber transport of multi-dimensional photonic qudits promises high information capacity per photon without space restriction. This work experimentally demonstrates transmission of spatial qudits through multi-core optical fibers and measurement of the entanglement between two fibers with quantum state analyzers, each composed of a spatial light modulator and a single-mode fiber. Quantum state tomography reconstructs the four-dimension entangled state that verifies the non-locality through concurrences in two-dimensional subspaces and a high-dimensional Bell-type CGLMP inequality.
\end{abstract}



\maketitle
High-dimensional encoding on single photons offers the potential to enhance quantum information processing. The dimensions added to simple qubits can increase the data rate~\cite{Walborn.PRL(2006)}, lower the acceptable error rate for secure quantum communications~\cite{Cerf.PRL(2002)}, and simplify quantum logic circuits~\cite{Lanyon.NP(2008)}. Spatial modes~\cite{Mair.Nature(2001), Dada.NP(2011), Dixon.PRL(2012), Malik.NC(2014), Krenn.PNAS(2014)}, temporal modes~\cite{Marcikic.PRL(2004), Hayat.OE(2012)} or colors~\cite{Ramelow.PRL(2009)} of photons can carry such multi-dimensional quantum states, namely qudits.

Spatial modes such as optical paths and orbital angular momentum states provide an advantage of relative ease in controlling arbitrary superpositions of multiple logical states $|1\rangle$, $|2\rangle$, $\cdots$, $|d\rangle$ ($d>2$). To avoid decoherence between the spatial modes due to ambient vibration noise, a practical implementation usually consists of paraxial modes sharing common holograms for state control~\cite{Mair.Nature(2001)} or inherently stable structures using beam displacing prisms and Sagnac interferometers~\cite{Lanyon.NP(2008), Gao.PRL(2010), Lee.OE(2012)}. This work shows that an optical fiber with multiple cores can also transport high-dimensional spatial quantum states. Fiber transport of spatial qubits ($d=2$) through a few-mode fiber has been demonstrated~\cite{Loffler.PRL(2011), Kang.PRL(2012)}, and we extend the work to a scheme that ideally can transmit large-$d$ qudits without inter-modal decoherence.

Multi-core fibers (MCFs) have been developed for high-power fiber laser amplifiers based on phase-locked beam combinations~\cite{Huo.OE(2004)} and more recently for space-division-multiplexing optical communications~\cite{Richardson.NP(2013), Uden.NP(2014)}. This work employs a commercially available four-core fiber to guide spatially-entangled telecom-wavelength photon pairs produced by spontaneous parametric down-conversion (SPDC). After transmission through the MCFs, each ququart ($d=4$) is analyzed through a spatial light modulator (SLM) and spatial-mode filtering by a single-mode fiber (SMF). The two-ququart entanglement is verified by concurrences in six pairwise two-mode subspaces and a generalized Bell inequality for high dimensions~\cite{Collins.PRL(2002)}, calculated from the result of quantum state tomography composed of 256 projection measurements.

\begin{figure}
\includegraphics[width=.45\textwidth]{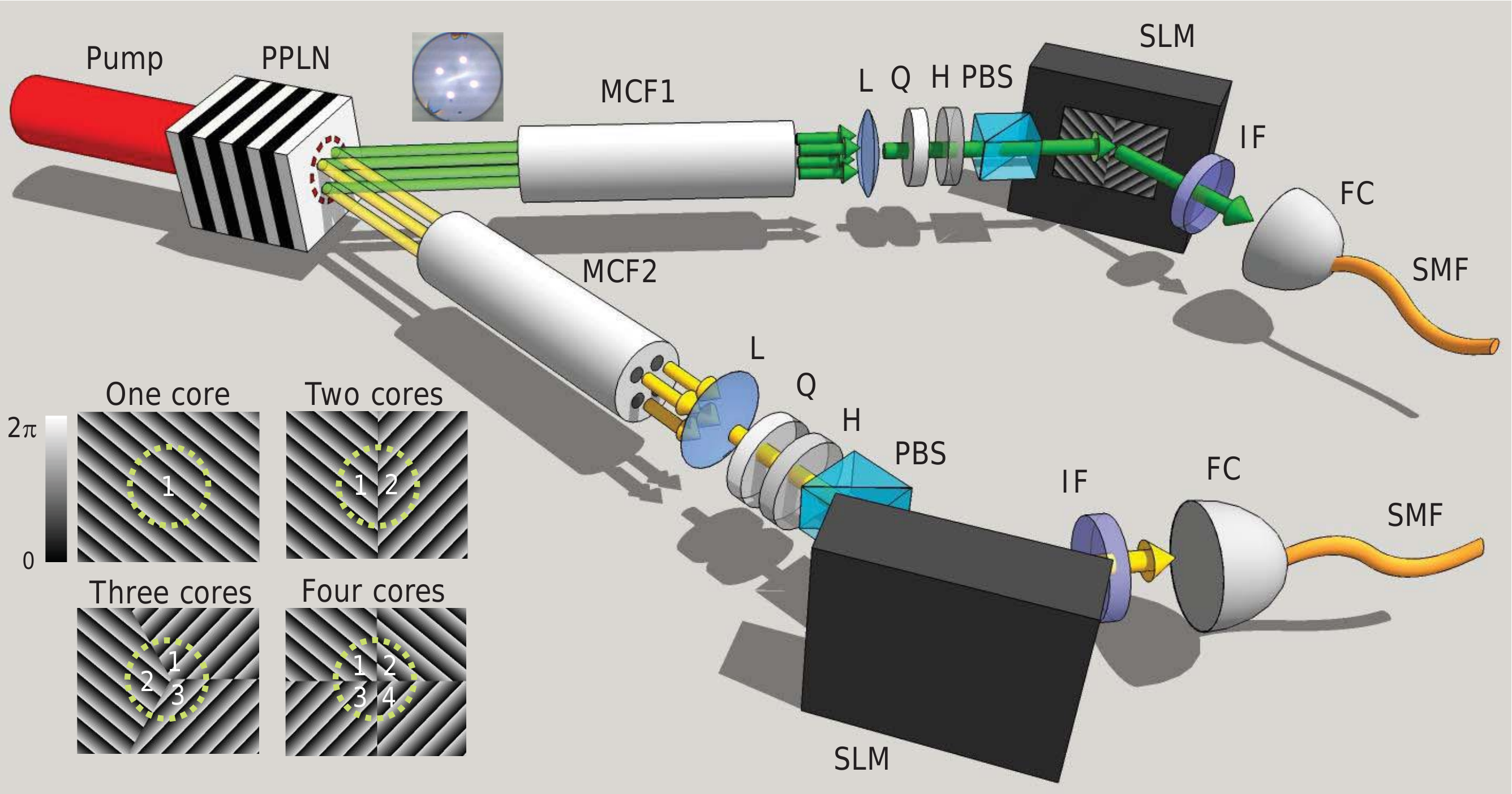}
\caption{Schematic of the experimental setup. Photon pairs are produced by non-collinear degenerate type-0 SPDC pumped by a continuous-wave laser beam with wavelength of 780~nm. (Imaging lenses between the PPLN crystal and the MCFs are omitted.) Insets: MCF cross-section and phase patterns on the SLM. PPLN: periodically poled lithium niobate; MCF: multi-core fiber; L: lens; Q: quarter-wave plate; H: half-wave plate; PBS: polarizing beam splitter; SLM: spatial light modulator; IF: interference filter; FC: fiber coupler; SMF: single-mode fiber.}
\label{f1}
\end{figure}

With our experimental setup (Fig.~1), spatially entangled photon pairs are generated by non-collinear degenerate type-0 SPDC in a periodically poled lithium niobate (PPLN) crystal (Covesion MSHG1550, poling period 19.5~$\mu$m, length 1~mm), pumped by a continuous-wave diode laser (wavelength 780~nm, output power 67~mW, linewidth $< 1$~MHz, $1/e^2$ spot diameter 415~$\mu$m). The down-converted photon pairs with a centre wavelength of 1560~nm propagate with a divergence angle of $3.1^\circ$. Each photon is respectively coupled to an MCF (Fibercore SM-4C 1500, total length 30~cm) that has four identical single-mode cores (mode field diameter 8~$\mu$m, NA $0.14-0.17$) at the vertices of a 37~$\mu$m $\times$ 37~$\mu$m square. The photon-collecting end face of each MCF is imaged onto the PPLN surface with five-fold magnification. The images of the two MCFs coincide on the crystal, and all the core images are well within the pump beam spot.

We post-select the photon pairs coupled to the cores of the two MCFs. Because the distances between the core images are greater than the transverse shift of photons inside the PPLN crystal, only four combinations between the identical-index cores $i=1,2,3,4$ of both MCF1 and MCF2 yield non-zero overlap integrals that lead to photon pair generation~\cite{Walborn.PR(2010)}. The post-selected quantum state propagating through the MCFs can be expressed as a four-dimensional entangled state:

\begin{equation}
|\Psi\rangle = a|1\rangle_{1}|1\rangle_{2} \\ + b|2\rangle_{1}|2\rangle_{2} + c|3\rangle_{1}|3\rangle_{2} + d|4\rangle_{1}|4\rangle_{2} ~,
\label{state}
\end{equation}
where $|i\rangle_{j}$ denotes the single-photon state in core $i$ of MCF$j$, and $a$, $b$, $c$, and $d$ are complex constants determined by the pump beam profile and the imaging configuration. The initially vertically polarized photons undergo birefringence in the MCFs because of their curvature and intrinsic stress. The output polarizations differ between the cores in our experiment; therefore, we convert the `average' polarization to horizontal polarization using a quarter-wave plate and a half-wave plate after collimation~\cite{Note.Polarization}. The vertical-polarization component is filtered out by a polarizing beam splitter. This core dependence of the birefringence can in principle be suppressed using polarization-maintaining MCFs~\cite{Stone.OL(2014), Ramirez.OE(2015)}. The polarization-filtered photons are reflected by an SLM before being narrow-band-filtered with an interference filter (half-maximum bandwidth 8.3~nm) and detected by an SMF-coupled avalanche photodiode single-photon counter. The phase patterns on the SLMs (inset of Fig.~1) determine the spatial quantum state measured by the single-photon counters.

The SLM is a reflection-mode phase modulator with a 792 $\times$ 600 array of 20~$\mu$m $\times$ 20~$\mu$m pixels (Hamamatsu LCOS-SLM). The grating-like patterns in Fig.~1 control the reflection angle of the incident beam by means of the direction and magnitude of the gradient of the linear phase. The pattern becomes a saw-tooth function because of the phase reset between $2\pi$ and $0$. The SLM surface is divided into subsections that connect different cores to the output SMF when the superposition states of multiple core modes are measured~\cite{Lee.OE(2015)}. The images of the MCF and the SMF are centered at the SLM patterns (Fig.~1) to equalize the coupling efficiencies of the four MCF cores to the output SMF following the procedures described in \cite{Lee.OE(2015)}. Coupling to $N$-core states reduces the total coupling efficiency to $1/N$ in our scheme; for example, projection of the output state to $|\psi_{out}\rangle$ to $|\psi_{meas}\rangle = (|1\rangle + \cdots + |N\rangle)/\sqrt{N}$ results in coincidence counts proportional to $|\langle\psi_{meas}|\psi_{out}\rangle |^2 / N$. To avoid the reduction of the detection efficiency in practical applications of such large-$N$ states, optical interferometric circuits for unitary transformation of the spatial modes will have to be developed possibly utilizing integrated-optic devices~\cite{Ding.ArXiv(2016)}. The coupling efficiencies from one core of the MCF to the output SMF are 100\%, 25\%, 11\%, and 6.3\% in an ideal case, and 54\%, 13\%, 4.8\%, and 3.6\% in our experiments for one-, two-, three-, and four-core superposition states, respectively. The differences are attributed to the reflectivity (90\%) of the SLMs and the efficiency of the mode coupling by two aspheric lenses (focal length 15 mm) between the MCF and the SMF.

We first measure the spatial correlation in Eq.~\eqref{state} by the coincidence counts between the core modes (Fig.~2). The unwanted correlations, i.e. all the correlations except the four dominant ones, sum to $2.0 \pm 0.3\%$ of the total of all counts. The error is the statistical uncertainty corresponding to $\pm$$\sqrt{\mbox{counts}}$ of the coincidence counts. Accidental coincidence counts were negligible in our experimental conditions.

\begin{figure}
\includegraphics[width=.4\textwidth]{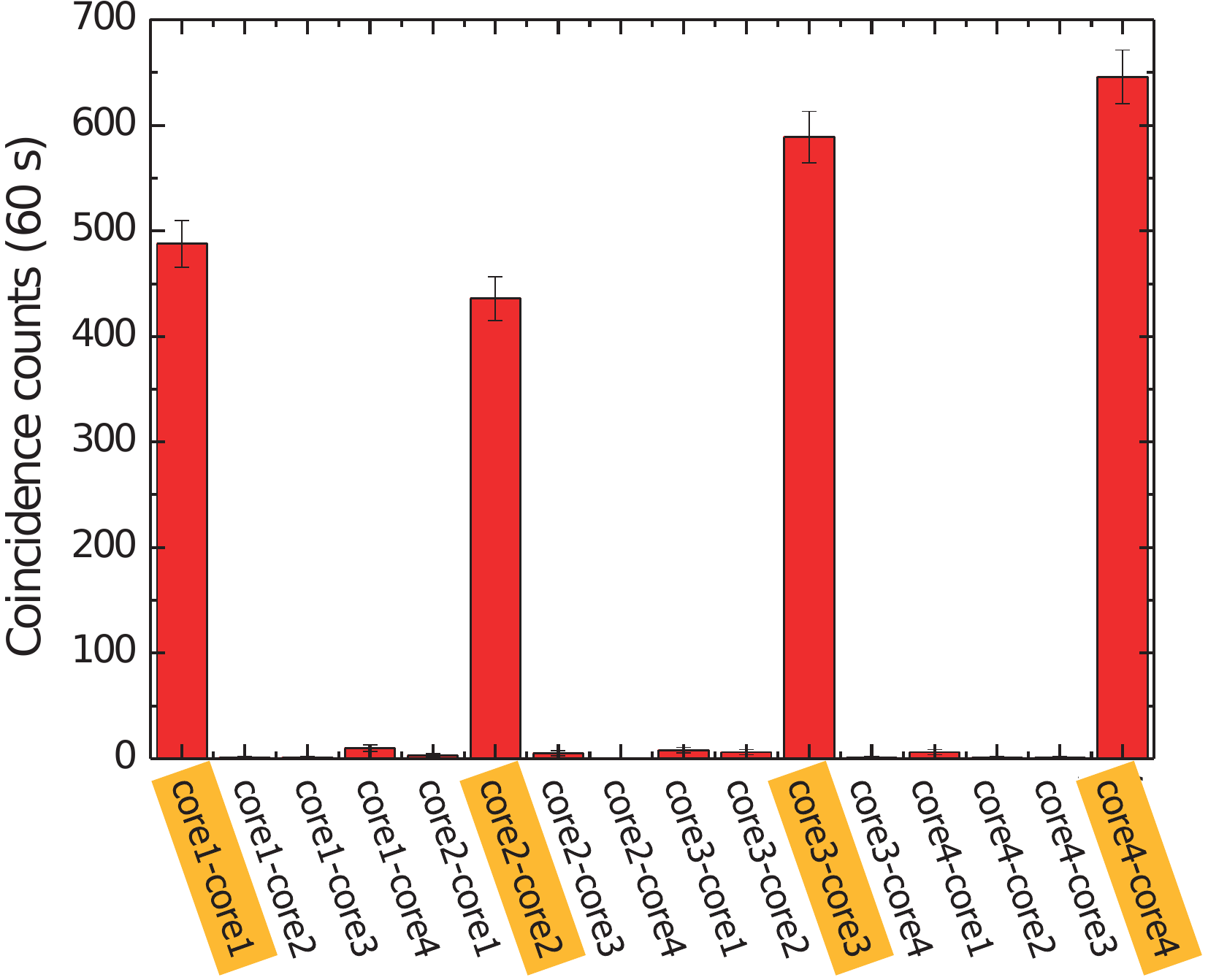}
\caption{Correlations between the core modes (core$i$-core$j$: projection to core $i$ of MCF1 and core $j$ of MCF2).}
\label{f2}
\end{figure}

Correlations between two-core superposition states verify the coherence between the four states in Eq.~\eqref{state} (Fig.~3). Patterns with two subsections are loaded on the SLM, where the relative phase $\phi_i (i=1,2)$ is adjusted by shifting the pattern of one subsection along the direction of the phase gradient. The results clearly show interference fringes with an average visibility of $0.88 \pm 0.02$. The phase biases of the interference fringes are attributed to a slight difference in refractive index between the cores of the MCF. We note that the SLMs do not introduce phase biases. Because the imaging configuration is symmetric, the optical path lengths from an MCF core to the output SMF are equal within 5\% of the wavelength even when reflected by different subsections of the SLM. Furthermore, we compensate for these small biases during the phase-sensitive measurements.

The sources of departure from the ideal visibility are under investigation. Core selectivity of the measurement setup does not significantly compromise the visibility because classical-light Mach-Zehnder interferometers using the same type of MCF and SLMs achieve visibilities greater than 99\%~\cite{Lee.OE(2015)}. Interference with the unwanted correlation components in Fig.~2 degrades the visibility by 0.02 on average. Non-uniformity of the four major components in Fig.~2 reduces the visibility by 0.01. Group refractive index mismatch between the cores of the MCF (within $6.5 \times 10^{-4}$)~\cite{Lee.OE(2015)} can also reduce the visibility when photons has finite wavelength bandwidth (8.3~nm) and the two MCFs has a length mismatch. Considering the uncertainty ($<$ 1~cm) in the fiber length measurement, the visibility reduction due to the last factor is $<0.003$. Remaining error sources cause the unexplained amount of decoherence between the terms in Eq.~\eqref{state}. Our non-collinear geometry can introduce slight temporal distinguishability between core modes by core-dependent path-length difference between the two MCFs. Non-uniformity of the MCF can also increase the decoherence due to the group index mismatch.

\begin{figure}
\includegraphics[width=.45\textwidth]{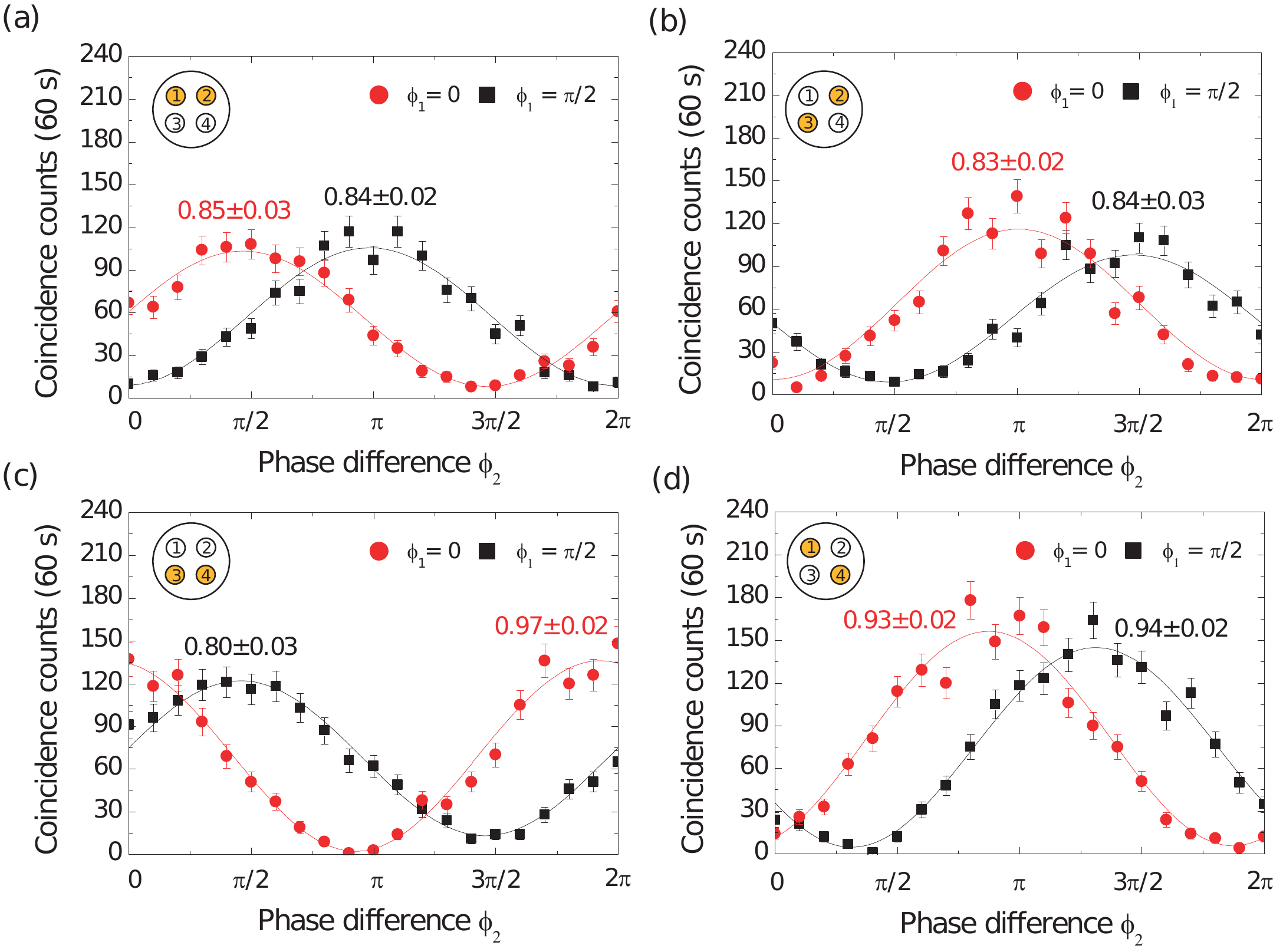}
\caption{Correlations between two-core superposition states. Photons 1 and 2 are projected onto the states $(|i\rangle_1 + e^{i\phi_{1}} |j\rangle_1 )/\sqrt{2}$ and $(|i\rangle_2 + e^{i\phi_{2}} |j\rangle_2 )/\sqrt{2}$, respectively. $\phi_1$ is fixed to $0$ (circles) or $\pi /2$ (squares). $\phi_2$ is scanned from $0$ to $2\pi$. The numbers are the fringe visibilities of the data. (a) $(i, j) = (1,2)$, (b) $(i, j) = (2,3)$, (c) $(i, j) = (3,4)$, (d) $(i, j) = (4,1)$ }
\label{f3}
\end{figure}

The interference fringe visibilities in Fig.~3 are a good measure for geometric positioning of the MCF cores. Because the cores in the MCFs are not perfectly identical, the same cores of MCF1 and MCF2 need to overlap to suppress the decoherence caused by the temporal distinguishability after propagation through the MCFs. With proper alignment, the coherence between the different core modes in Eq.~(1) is maintained if the lengths of the fibers are the same and the differential time delay between the cores is smaller than the coherence time ($>1 ~\mu\rm{s}$) of the pump laser. Otherwise maintaining the coherence of the entangled state requires the differential group delay to be smaller than the coherence time of the photons, that is, much shorter than $\sim 400 ~\rm{fs}$. Figure 4 shows the correlation between superposition states of two diagonal cores with MCF1 rotated by $0^\circ$, $90^\circ$, $180^\circ$, and $270^\circ$ and all other optical components fixed. We removed the interference filters during these measurements to better clarify the optimal position of the cores [Fig.~4(a)]. The non-ideal maximum visibility of $0.56 \pm 0.02$ may have resulted from a fiber length mismatch of $<1$~cm between MCF1 and MCF2 when considering the group index difference of $6.5 \times 10^{-4}$ between cores 1 and 4 and a half-maximum wavelength bandwidth greater than 150~nm for the photons.

\begin{figure}
\includegraphics[width=.45\textwidth]{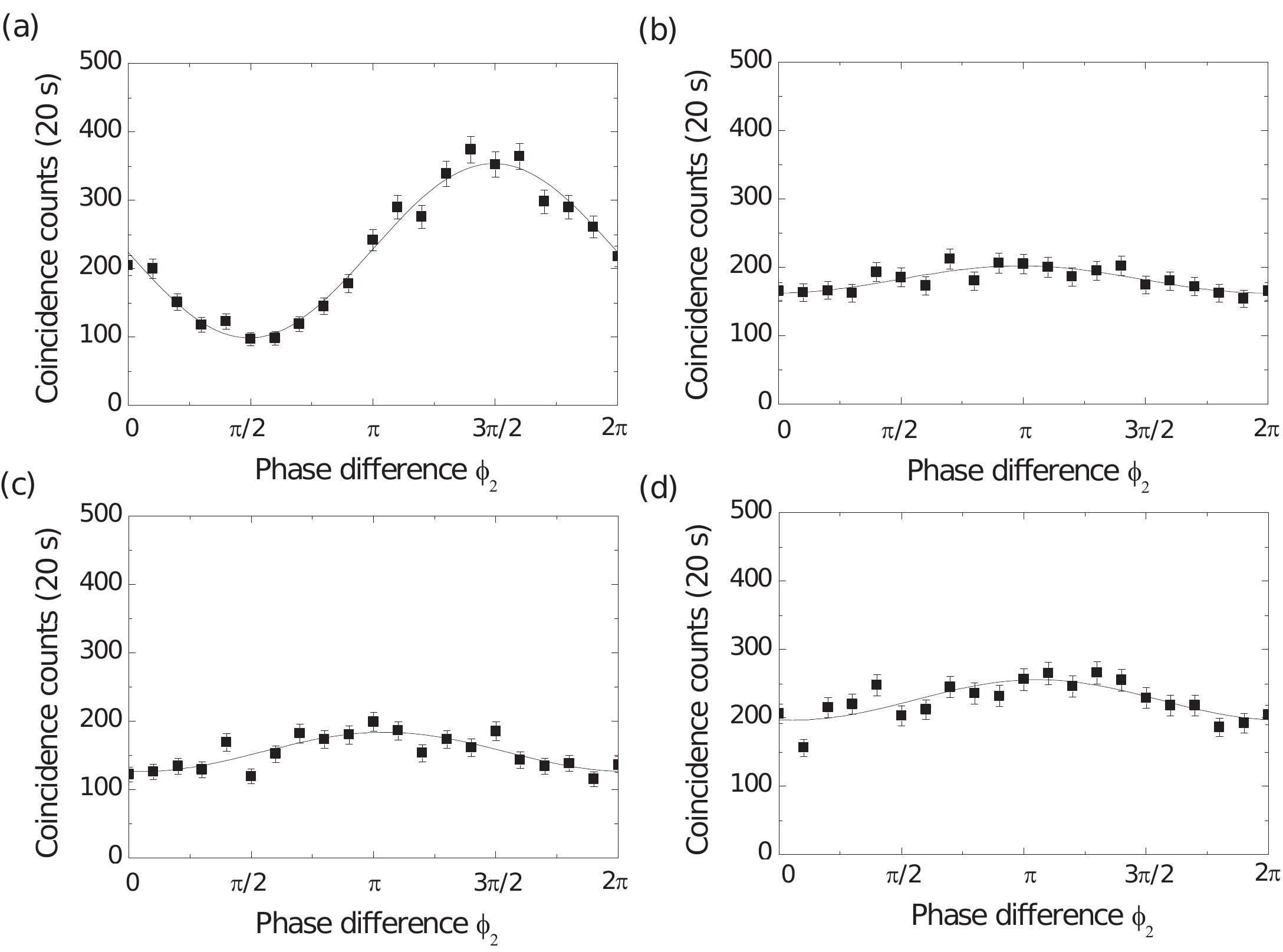}
\caption{Interference fringes with varying alignment of the four cores: Coincidence counts with photons 1 and 2 projected onto $(|1\rangle_1 + |4\rangle_1 )/\sqrt{2}$ and $(|1\rangle_2 + e^{i\phi_2}|4\rangle_2 )/\sqrt{2}$, respectively. MCF1 is rotated with respect to the fiber axis by (a) $0^\circ$, (b) $90^\circ$, (c) $180^\circ$, and (d) $270^\circ$. The interference filters in Fig.~1 are removed for these measurements.}
\label{f4}
\end{figure}

Quantum state tomography reconstructs the density matrix (Fig.~5) of the output state from $256=16\times16$ projection measurements composed of four one-core states and twelve two-core superposition states ($(|1\rangle + |2\rangle )/\sqrt{2})$, $(|1\rangle + i|2\rangle )/\sqrt{2})$, etc.) of each photon. Each projection measurement was coincidence counts for $60$~s. To better compare the result with the symmetric maximally entangled state $|\beta\rangle = 1/2(|1\rangle|1\rangle + |2\rangle|2\rangle + |3\rangle|3\rangle + |4\rangle|4\rangle)$, we re-phased the reconstructed state $\hat{\rho}_0$ by adding phases $0$, $\phi_2$, $\phi_3$, and $\phi_4$ to the $|1\rangle$, $|2\rangle$, $|3\rangle$, and $|4\rangle$ states, respectively, of photon 1. Here, $\phi_2 = \langle1|\langle1| \hat{\rho}_0 |2\rangle|2\rangle$, $\phi_3 = \langle1|\langle1| \hat{\rho}_0 |3\rangle|3\rangle$, and $\phi_4 = \langle1|\langle1| \hat{\rho}_0 |4\rangle|4\rangle$. Hence the final density matrix $\hat{\rho}$ (Fig.~5) is given by $\hat{\rho} = U \hat{\rho}_0 U^\dag$ ($U = M\otimes I$), where $M$ is the phase shift operator and $I$ is the identity operator. Ideally the sixteen peaks in Fig.~5(a) should be constant at 0.25 and the other elements in Fig.~5(a) and (b) should be zero.

The Schmidt number of the reconstructed state is $3.79 \pm 0.03$, which verifies the dimensionality of the state. The fidelity $\rm{Tr}\{\sqrt{\hat{\rho}^{1/2} |\beta\rangle \langle \beta | \hat{\rho}^{1/2}}\}$ with the ideal maximally entangled state is $0.91 \pm 0.01$, and the state purity $\rm{Tr}\{ \hat{\rho}^2\}$ is $0.73 \pm 0.03$. To quantify the quantum coherence between the components, we calculate the concurrences of the two-dimensional subspaces (Table 1).

\begin{figure}
\includegraphics[width=.4\textwidth]{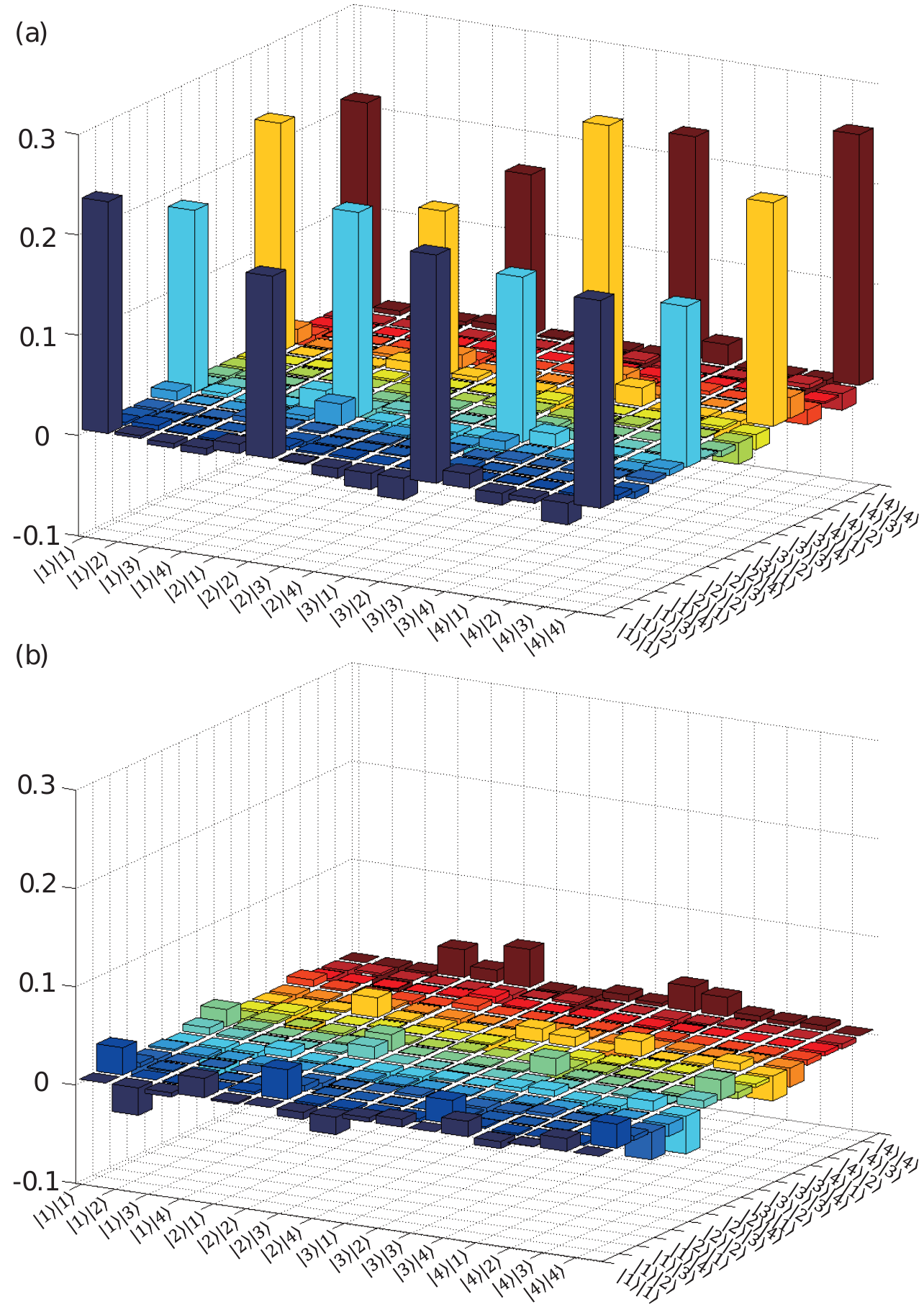}
\caption{Two-ququart density matrix $\hat{\rho}$ reconstructed by quantum state tomography: (a) real and (b) imaginary parts.}
\label{f5}
\end{figure}

\begin{table}[b]
\caption{\label{tab:table1} Concurrences of the two-dimensional subspaces.}
\begin{ruledtabular}
\begin{tabular}{lcdr}
\multicolumn{1}{c}{Ideal State}&\multicolumn{1}{c}{Experimental Concurrence}\\
\colrule
1/$\sqrt{2}(|1\rangle_{1}|1\rangle_{2}+|2\rangle_{1}|2\rangle_{2}$) & $0.82 \pm 0.05$\\
1/$\sqrt{2}(|1\rangle_{1}|1\rangle_{2}+|3\rangle_{1}|3\rangle_{2}$) & $0.89 \pm 0.04$\\
1/$\sqrt{2}(|1\rangle_{1}|1\rangle_{2}+|4\rangle_{1}|4\rangle_{2}$) & $0.85 \pm 0.03$\\
1/$\sqrt{2}(|2\rangle_{1}|2\rangle_{2}+|3\rangle_{1}|3\rangle_{2}$) & $0.67 \pm 0.06$\\
1/$\sqrt{2}(|2\rangle_{1}|2\rangle_{2}+|4\rangle_{1}|4\rangle_{2}$) & $0.72 \pm 0.06$\\
1/$\sqrt{2}(|3\rangle_{1}|3\rangle_{2}+|4\rangle_{1}|4\rangle_{2}$) & $0.85 \pm 0.04$\\
\end{tabular}
\end{ruledtabular}
\end{table}
We further test the generalized Bell-type CGLMP inequality for a high-dimensional two-photon quantum state~\cite{Collins.PRL(2002), Chen.PRA(2006)} with the reconstructed state. The Bell parameter $I_{d}$ $(d\geq2)$ fulfils the inequality $I_{d} \leq 2$ by local variable theories, and reaches its maximum of $I_4 = 2.9727$ with a non-maximally entangled state~\cite{Chen.PRA(2006)}. The symmetric maximally entangled state theoretically yields a value of $I_{4} = 2.8962$. We use the general Bell operator matrix form~\cite{Chen.PRA(2006)} to calculate $I_4$ from the density matrix (Fig.~5), which leads to $I_4 = 2.27 \pm 0.06$ violating the inequality by 4.5 standard deviations.

We discuss an extension of our scheme to general entangled states and to longer-distance transmission. Coefficients $a$, $b$, $c$, and $d$ in Eq.~\eqref{state} depend on the spatial profile of the pump beam. The same technique using an SLM to control the photonic spatial mode is also applicable to the pump laser to realize an arbitrary combination of the coefficients. Quantum information processing based on hyper-entanglement or hybrid entanglement can be realized if one incorporates qubit-joining or qubit-transduction techniques that can convert multiple qubits into a single spatial qudit~\cite{Vitelli.NP(2013), Lim.SR(2015), Feng.NC(2016)}. Ideal MCFs can transport arbitrary quantum states over long distances without suffering decoherence. However, slight differences in group refractive indices between the cores of a practical MCF limit the transmission distance, which is less than a few meters with the MCF in our experiments. This limit can be overcome by placing mode converters that cyclically permutate the core modes, core $(1,2,\dots,d)$ to core $(2,3,\dots,d,1)$, at each of the $d-1$ uniformly distributed points. When the cores are located on the circumference of a circle as in Fig.~1, this mode converter can be realized by splicing two MCF sections with a relative angle detuning of $360/d^\circ$.

In summary, we have proven the experimental feasibility of transmission of four-dimensional spatially entangled photon pairs through commercially available multi-core optical fibers. Quantum state tomography composed of one-core states and two-core superposition states has quantitatively verified the non-classicality of the transmitted photonic states through the concurrences of six two-dimensional subspaces and the violation of the four-dimensional CGLMP inequality. These results suggest the usefulness of optical fibers for multi-dimensional quantum communications and quantum interfaces.

\begin{acknowledgments}
The authors thank Jae Yong Lee, Jae Bong Song and YongKeun Park for help with the experimental setup, and I. R. Berchera for discussion. This work was supported by the R\&D Convergence Program of the NST (Grant No. CAP-15-08-KRISS), the International Research \& Development Program of the NRF (Grant No. NRF-2015K1A3A7A03074107), and the KRISS project `Convergent Science and Technology for Measurement at the Nanoscale' of Republic of Korea.
\end{acknowledgments}


\begin{thebibliography}{29}
\bibitem [{Not()}]{Note.Polarization}%
  \BibitemOpen
  \href@noop {} {}\bibinfo {note} {Polarization-state tomography showed that
  the fidelities (the inner-product-squared of two Jones vectors) between the
  output polarizations of different cores ranged 0.07-0.75 and 0.52-0.97 for
  MCF1 and MCF2, respectively. The average polarization was calculated as a
  normalized average of the polarization vectors on the Poincar\'{e}
  sphere.}\BibitemShut {Stop}%
\end{thebibliography}
\end{document}